\documentclass[stslayout, noinfoline]{imsart}
\usepackage{geometry} 
\usepackage{graphicx}	
\usepackage{amssymb, amsmath}
\usepackage{natbib}
\usepackage[colorlinks,citecolor=blue,urlcolor=blue]{hyperref}

\usepackage{tikz}
\usetikzlibrary{intersections, backgrounds}
\usepackage{pgfmath}

\definecolor{light}{RGB}{220, 188, 188}
\definecolor{mid}{RGB}{185, 124, 124}
\definecolor{dark}{RGB}{143, 39, 39}
\definecolor{highlight}{RGB}{180, 31, 180}
\definecolor{gray10}{gray}{0.1}
\definecolor{gray20}{gray}{0.2}
\definecolor{gray30}{gray}{0.3}
\definecolor{gray40}{gray}{0.4}
\definecolor{gray60}{gray}{0.6}
\definecolor{gray70}{gray}{0.7}
\definecolor{gray80}{gray}{0.8}
\definecolor{gray90}{gray}{0.9}
\definecolor{gray95}{gray}{0.95}

\begin{document}

\begin{frontmatter}

\title{The Discrete Adjoint Method: Efficient Derivatives for\\Functions of Discrete Sequences}
\runtitle{The Discrete Adjoint Method}

\begin{aug}
  \author{Michael Betancourt%
  \ead[label=e1]{betan@symplectomorphic.com}},
  \author{Charles C. Margossian},
  \author{Vianey Leos-Barajas}
  
  \runauthor{Betancourt et al.}

  \address{Michael Betancourt is the principal research scientist
           at Symplectomorphic, LLC. \printead{e1}.  Charles Margossian
           is a PhD candidate in the Department of Statistics, 
           Columbia University.  Vianey Leos-Barajas is a postdoctoral
           researcher in the Department of Forestry and Environmental 
           Resources and the Department of Statistics at North Carolina 
           State University.}
\end{aug}

\begin{abstract}
Gradient-based techniques are becoming increasingly critical in quantitative fields, 
notably in statistics and computer science.  The utility of these techniques, however, 
ultimately depends on how efficiently we can evaluate the derivatives of the complex 
mathematical functions that arise in applications.  In this paper we introduce a 
discrete adjoint method that efficiently evaluates derivatives for functions of discrete 
sequences.
\end{abstract}

\end{frontmatter}

Many popular mathematical models, such as common hidden Markov models, utilize 
sequences of discrete states implicitly defined through \emph{forward difference equations},
\begin{equation*}
\mathbf{u}_{n + 1} - \mathbf{u}_{n} = \boldsymbol{\Delta}_{n} (\mathbf{u}_{n}, \psi, n),
\end{equation*}
to capture the regular evolution of a latent system; here $\mathbf{u}_{n}$ denotes 
the $n$th latent state of the system and $\psi$ the model parameters. Typically these 
sequences are incorporated into larger models through \emph{discrete functionals}
that consume particular sequences and return scalar values,
\begin{equation*}
\mathcal{J}(\psi) = \sum_{n = 0}^{N - 1} j_{n}(\mathbf{u}_{n}, \psi, n).
\end{equation*}

We can quantify the impact of the parameters, $\psi$, on these functionals
by evaluating the total derivatives, $\mathrm{d} \mathcal{J} / \mathrm{d} \psi$.  
The evaluation of these derivatives is complicated by the dependence of the sequences 
on the parameters enforced by the forward difference equations; the total derivative 
of a functional has to take into account both the \emph{explicit} dependence of the 
$j_{n}$ on $\psi$ and also the \emph{implicit} dependence mediated by the latent states 
$\mathbf{u}_{n}$.

We can always compute each \emph{sensitivity}, $\mathrm{d} \mathbf{u}_{n} / \mathrm{d} \psi$,
by propagating derivatives along the forward difference equations and constructing the
corresponding sequence of sensitivities. This quickly becomes expensive, however, 
when there are many parameters that each require their own sensitivities.
In order to better scale we need to bypass the superfluous computation of these
intermediate derivatives and only propagate the minimal information
needed to construct the total derivatives of the desired functionals.

In this paper we introduce a \emph{discrete adjoint} technique that efficiently 
computes total derivatives without explicitly calculating intermediate sensitivities.
We begin by reviewing the powerful continuous adjoint method for ordinary differential 
equations before deriving a discrete analog.  Finally we demonstrate how the method
can be applied to hidden Markov models.

\section{Continuous Adjoint Systems}

The continuous analog of discrete sequences are \emph{state trajectories}, $\mathbf{u}(t)$, 
defined implicitly through the ordinary differential equations
\begin{equation*}
\frac{ \mathrm{d} \mathbf{u} }{ \mathrm{d} t } = \mathbf{f}(\mathbf{u}, \psi, t)
\end{equation*}
along with the initial conditions
\begin{equation*}
\mathbf{u}(t = 0) = \boldsymbol{\upsilon}(\psi).
\end{equation*}
A \emph{functional} consumes the state trajectory and returns a single real number 
through an integration over time,
\begin{equation*}
\mathcal{J}(\psi) = \int_{0}^{T} \mathrm{d}t \, j(\mathbf{u}, \psi, t).
\end{equation*}
Our goal is then to compute the \emph{total} derivative of $\mathcal{J}$ with 
respect to the parameter $\psi$, taking into account not only the explicit dependence 
of $\psi$ on $j$ but also the implicit dependence through the influence of $\psi$ on 
the evolution of the states $\mathbf{u}(t)$.  For a thorough review of the possible
strategies see Section 2.6 and 2.7 of \cite{HindmarshEtAl:2020}.

\subsection{Adjoint Task Force}

An immediate way to compute gradients of functionals like this is to explicitly compute 
the state sensitivities
\begin{equation*}
\boldsymbol{\eta} = \mathrm{d} \mathbf{u} / \mathrm{d} \mathrm{\psi}
\end{equation*}
by solving the auxiliary ordinary differential equations,
\begin{align*}
\frac{ \mathrm{d} \boldsymbol{\eta} }{ \mathrm{d} t } 
&=
\frac{ \mathrm{d} }{ \mathrm{d} t } 
\left( \frac{ \mathrm{d} \mathbf{u} }{ \mathrm{d} \psi } \right)
\\
&=
\frac{ \mathrm{d} }{ \mathrm{d} \psi } 
\left( \frac{ \mathrm{d} \mathbf{u} }{ \mathrm{d} t } \right)
\\
&=
\frac{ \mathrm{d} }{ \mathrm{d} \psi } 
\bigg( \mathbf{f} \bigg)
\\
&=
\frac{ \partial \mathbf{f} }{ \partial \psi}
+ 
\left( \frac{ \mathrm{d} \mathbf{f} }{ \mathrm{d} \mathbf{u} } \right)^{\dagger}
\cdot \frac{ \mathrm{d} \mathbf{u} }{ \mathrm{d} \psi } 
\\
&=
\frac{ \partial \mathbf{f} }{ \partial \psi}
+ 
\left( \frac{ \mathrm{d} \mathbf{f} }{ \mathrm{d} \mathbf{u} } \right)^{\dagger}
\cdot \boldsymbol{\eta},
\end{align*}
Here a boldfaced fraction is shorthand for the Jacobian matrix
\begin{equation*}
\left( \frac{ \mathrm{d} \mathbf{f} }{ \mathrm{d} \mathbf{u} } \right)_{ij}
=
\frac{ \mathrm{d} f_{i} }{ \mathrm{d} u_{j} }.
\end{equation*}

Once we've solved for the state sensitivities we can construct the total 
derivative of the desired functional through the chain rule,
\begin{align*}
\frac{ \mathrm{d} \mathcal{J} }{ \mathrm{d} \psi} (\psi)
&=
\frac{ \mathrm{d} }{ \mathrm{d} \psi }
\int_{0}^{T} \mathrm{d} t \, j(\mathbf{u}, \psi, t)
\\
&=
\int_{0}^{T} \mathrm{d} t \, 
\frac{ \mathrm{d} }{ \mathrm{d} \psi } j(\mathbf{u}, \psi, t)
\\
&=
\int_{0}^{T} \mathrm{d} t \left[
\frac{ \partial j}{ \partial \psi }
+ \left( \frac{ \mathrm{d} j }{ \mathrm{d} \mathbf{u} } \right)^{\dagger}
\cdot \boldsymbol{\eta} \right].
\end{align*}
This approach becomes burdensome, however, once we consider multiple parameters 
and hence multiple total derivatives, each of which requires integrating over 
its own trajectory of sensitivities.

Another way to work out the total derivative of the functional is to treat
the influence of the parameter on the state trajectory as \emph{constraints}
\citep{HannemannEtAl:2015},
\begin{align*}
0 &= \mathbf{u}(0) - \boldsymbol{\upsilon}(\psi)
\\
0 &= \frac{ \mathrm{d} \mathbf{u} }{ \mathrm{d} t } 
- \mathbf{f}(\mathbf{u}, \psi, t),
\end{align*}
which are explicitly incorporated into the functional with \emph{Lagrange 
multipliers}, $\boldsymbol{\mu}$ and $\boldsymbol{\lambda}(t)$,
\begin{align*}
\mathcal{J}(\psi) 
&= \int_{0}^{T} \mathrm{d}t \, j(\mathbf{u}, \psi, t)
\\
&= 0 + \int_{0}^{T} \mathrm{d}t \, j(\mathbf{u}, \psi, t) + 0
\\
&= \boldsymbol{\mu}^{\dagger} \cdot \left[ \mathbf{u}(0) - \boldsymbol{\upsilon}(\psi) \right]
+ \int_{0}^{T} \mathrm{d}t \, j(\mathbf{u}, \psi, t)
+ \boldsymbol{\lambda}^{\dagger}(t) \cdot \left[ \frac{ \mathrm{d} \mathbf{u} }{ \mathrm{d} t } 
- \mathbf{f}(\mathbf{u}, \psi, t) \right]
\\
&\equiv \mathcal{L}(\psi).
\end{align*}
As long as the constraints are satisfied this modified functional will
equal our target functional for \emph{any} values of the Lagrange multipliers.

Under these constraints we can compute the total derivative of the functional 
by instead differentiating this modified functional.  If we assume that everything
is smooth then we can exchange the order of integration and differentiation to give
\begin{align*}
\frac{ \mathrm{d} \mathcal{J} }{ \mathrm{d} \psi}
&=
\frac{ \mathrm{d} \mathcal{L} }{ \mathrm{d} \psi}
\\
&=
\boldsymbol{\mu}^{\dagger} \cdot \left[ 
\frac{ \mathrm{d} \mathbf{u}}{ \mathrm{d} \psi}(0) 
- \frac{ \mathrm{d} \boldsymbol{\upsilon}}{ \mathrm{d} \psi}
\right]
+ \int_{0}^{T} \mathrm{d}t \, 
\frac{ \mathrm{d} j }{ \mathrm{d} \psi}
+ \boldsymbol{\lambda}^{\dagger}(t) \cdot \left[ 
\frac{ \mathrm{d} }{ \mathrm{d} \mathrm{\psi} }
\frac{ \mathrm{d} \mathbf{u} }{ \mathrm{d} t } 
- 
\frac{ \mathrm{d} \mathbf{f} }{ \mathrm{d} \mathrm{\psi} } \right]
\\
&=
\boldsymbol{\mu}^{\dagger} \cdot \left[ 
\frac{ \mathrm{d} \mathbf{u}}{ \mathrm{d} \psi}(0) 
- \frac{ \partial \boldsymbol{\upsilon}}{ \partial \psi}
\right]
+ \int_{0}^{T} \mathrm{d}t \, 
 \left[\frac{ \partial j }{ \partial \psi}
+ \left( \frac{ \partial j }{ \partial \mathbf{u} } \right)^{\dagger}
\cdot \frac{ \mathrm{d} \mathbf{u} }{ \mathrm{d} \psi} \right]
+ \boldsymbol{\lambda}^{\dagger}(t) \cdot \left[ 
\frac{ \mathrm{d} }{ \mathrm{d} t }
\frac{ \mathrm{d} \mathbf{u} }{ \mathrm{d} \mathrm{\psi} } 
- 
\frac{ \partial \mathbf{f} }{ \partial \mathrm{\psi} } 
- \left( \frac{ \partial \mathbf{f} }{ \partial \mathbf{u} } \right)^{\dagger}
\cdot \frac{ \mathrm{d} \mathbf{u} }{ \mathrm{d} \psi } 
\right].
\end{align*}
Once again a boldfaced fraction is shorthand for a Jacobian matrix.  For example,
\begin{equation*}
\frac{ \partial j }{ \partial \mathbf{u} }
=
\left( \frac{ \partial j }{ \partial u_{1} }, 
\ldots, \frac{ \partial j }{ \partial u_{N} } \right)^{\dagger}.
\end{equation*}

The benefit of this approach is that we can use the freedom in our Lagrange multipliers 
to eliminate the expensive state sensitivities entirely!  First we need to integrate the 
time derivative of the sensitivities by parts to recover a pure sensitivity,
\begin{equation*}
\int_{0}^{T} \mathrm{d}t \, 
\boldsymbol{\lambda}^{\dagger}(t) \cdot \frac{ \mathrm{d} }{ \mathrm{d} t }
\frac{ \mathrm{d} \mathbf{u} }{ \mathrm{d} \mathrm{\psi} } 
=
\boldsymbol{\lambda}^{\dagger}(T) \cdot \frac{ \mathrm{d} \mathbf{u} }{ \mathrm{d} \mathrm{\psi} } (T)
- \boldsymbol{\lambda}^{\dagger}(0) \cdot \frac{ \mathrm{d} \mathbf{u} }{ \mathrm{d} \mathrm{\psi} } (0)
- \int_{0}^{T} \mathrm{d}t \, 
\left( \frac{ \mathrm{d} \boldsymbol{\lambda} }{ \mathrm{d} t} \right)^{\dagger} \cdot
\frac{ \mathrm{d} \mathbf{u} }{ \mathrm{d} \mathrm{\psi} }.
\end{equation*}
Then we substitute this result into the total derivative and gather all the sensitivity
terms together,
\begin{align*}
\frac{ \mathrm{d} \mathcal{J} }{ \mathrm{d} \psi}
&=
\quad
\boldsymbol{\mu}^{\dagger} \cdot \left[ 
\frac{ \mathrm{d} \mathbf{u}}{ \mathrm{d} \psi}(0) 
- \frac{ \partial \boldsymbol{\upsilon}}{ \partial \psi}
\right]
+ \boldsymbol{\lambda}^{\dagger}(T) \cdot \frac{ \mathrm{d} \mathbf{u} }{ \mathrm{d} \mathrm{\psi} } (T)
- \boldsymbol{\lambda}^{\dagger}(0) \cdot \frac{ \mathrm{d} \mathbf{u} }{ \mathrm{d} \mathrm{\psi} } (0)
\\
&\quad+
\int_{0}^{T} \mathrm{d}t \, 
\frac{ \partial j }{ \partial \psi}
+ \left( \frac{ \partial j }{ \partial \mathbf{u} } \right)^{\dagger}
\cdot \frac{ \mathrm{d} \mathbf{u} }{ \mathrm{d} \psi}
- \left( \frac{ \mathrm{d} \boldsymbol{\lambda} }{ \mathrm{d} t} \right)^{\dagger} \cdot
\frac{ \mathrm{d} \mathbf{u} }{ \mathrm{d} \mathrm{\psi} }
- \boldsymbol{\lambda}^{\dagger}(t) \cdot
\frac{ \partial \mathbf{f} }{ \partial \mathrm{\psi} }
- \boldsymbol{\lambda}^{\dagger}(t) \cdot \left( \frac{ \partial \mathbf{f} }{ \partial \mathbf{u} } \right)^{\dagger}
\cdot \frac{ \mathrm{d} \mathbf{u} }{ \mathrm{d} \psi }
\\
&=
\quad
\bigg[ \boldsymbol{\mu} - \boldsymbol{\lambda}(0) \bigg]^{\dagger} 
\cdot \frac{ \mathrm{d} \mathbf{u}}{ \mathrm{d} \psi}(0) 
- \boldsymbol{\mu}^{\dagger} \cdot \frac{ \partial \boldsymbol{\upsilon}}{ \partial \psi}
+ \boldsymbol{\lambda}^{\dagger}(T) \cdot \frac{ \mathrm{d} \mathbf{u} }{ \mathrm{d} \mathrm{\psi} } (T)
\\
&\quad+
\int_{0}^{T} \mathrm{d}t \, 
\frac{ \partial j }{ \partial \psi}
- \boldsymbol{\lambda}^{\dagger}(t) \cdot
\frac{ \partial \mathbf{f} }{ \partial \mathrm{\psi} } 
+ \int_{0}^{T} \mathrm{d}t \, 
\left[ \frac{ \partial j }{ \partial \mathbf{u} }
- \frac{ \mathrm{d} \boldsymbol{\lambda} }{ \mathrm{d} t}
- \boldsymbol{\lambda}(t) \cdot \frac{ \partial \mathbf{f} }{ \partial \mathbf{u} } 
\right]^{\dagger} \cdot \frac{ \mathrm{d} \mathbf{u} }{ \mathrm{d} \psi}
\end{align*}

Now we can exploit the freedom in our Lagrange multipliers to remove all vestiges of the 
sensitivities.  First let's set $\boldsymbol{\mu} = \boldsymbol{\lambda}(0)$ to remove the
initial sensitivities and $\boldsymbol{\lambda}(T) = 0$ to remove the final sensitivities.
We can then remove the integral term that depends on the intermediate sensitivities if we set
\begin{equation*}
\frac{ \partial j }{ \partial \mathbf{u} }
- \frac{ \mathrm{d} \boldsymbol{\lambda} }{ \mathrm{d} t}
- \boldsymbol{\lambda}(t) \cdot \frac{ \partial \mathbf{f} }{ \partial \mathbf{u} } 
= 0,
\end{equation*}
or
\begin{equation*}
\frac{ \mathrm{d} \boldsymbol{\lambda} }{ \mathrm{d} t}
=
\frac{ \partial j }{ \partial \mathbf{u} }
- \boldsymbol{\lambda}(t) \cdot \frac{ \partial \mathbf{f} }{ \partial \mathbf{u} }.
\end{equation*}

In other words provided that $\boldsymbol{\lambda}(t)$ satisfies the differential
equation
\begin{equation*}
\frac{ \mathrm{d} \boldsymbol{\lambda} }{ \mathrm{d} t}
=
\frac{ \partial j }{ \partial \mathbf{u} }(\mathbf{u}, \psi, t)
- \boldsymbol{\lambda}(t) \cdot 
\frac{ \partial \mathbf{f} }{ \partial \mathbf{u} }(\mathbf{u}, \psi, t)
\end{equation*}
with the initial conditions
\begin{equation*}
\boldsymbol{\lambda}(T) = 0
\end{equation*}
then then total derivative of our target functional reduces to
\begin{equation*}
\frac{ \mathrm{d} \mathcal{J} }{ \mathrm{d} \psi}(\psi)
=
- \boldsymbol{\lambda}^{\dagger}(0) \cdot \frac{ \partial \boldsymbol{\upsilon}}{ \partial \psi}
+ \int_{0}^{T} \mathrm{d}t \, 
\frac{ \partial j }{ \partial \psi}(\mathbf{u}, \psi, t)
- \boldsymbol{\lambda}^{\dagger}(t) \cdot
\frac{ \partial \mathbf{f} }{ \partial \mathrm{\psi} }(\mathbf{u}, \psi, t).
\end{equation*}
The system of differential equations for $\boldsymbol{\lambda}(t)$ is known as the 
\emph{adjoint} system relative to the original system of ordinary differential 
equations.  If we first solve for $\mathbf{u}(t)$ then we can solve for the adjoint 
$\boldsymbol{\lambda}(t)$ and compute the total derivative 
$\mathrm{d} \mathcal{J} / \mathrm{d} \psi$ at the same time without having to
compute any explicit sensitivities.

\subsection{Computational Scalings}

For a single parameter the direct approach is slightly more efficient, requiring two 
$N$-dimensional integrations for the states and their sensitivities compared to the 
adjoint approach which requires two $N$-dimensional integrations, one for the states 
and one for the adjoint states, \emph{and} the extra one-dimensional integration to 
solve for the total derivative.  The adjoint method, however, quickly becomes more 
efficient as we consider multiple parameters because the adjoint states 
\emph{are the same for all parameters}.  

When we have $K$ parameters the forward sensitivity approach requires an $N$-dimensional 
integration for \emph{each} sensitivity and the total cost scales as $N + N \cdot K$.
The adjoint approach, however, requires only two $N$-dimensional solves to set up the
states and the adjoint states and then $K$ one-dimensional solves for each gradient 
component, yielding a total cost scaling of $2 N + K$.

Comparing these two scalings we see that the adjoint method is better when
\begin{equation*}
  \frac{N}{N - 1} < K,
\end{equation*}
a condition verified for any $N$ provided that $K \ge 2$.  In other words the adjoint
method will generally feature the highest performance in any application with at least 
two parameters.  As the number of parameters increases the $\mathcal{O}(NK)$ scaling of 
the forward sensitivity approach grows much faster than the $\mathcal O (N + K)$ scaling
of the adjoint method, and the performance gap only becomes more substantial.

\subsection{An Application to Automatic Differentiation}

A particularly useful application of the continuous adjoint method is for the reverse 
mode automatic differentiation \citep{BuckerEtAl:2006, GriewankEtAl:2008, Margossian:2019}
of functions incorporating the solutions of ordinary differential equations.  In order 
to propagate the needed differential information through the composite function we need 
to be able to evaluate the Jacobian of the final state with respect to the parameters, 
\begin{equation*}
\frac{ \mathrm{d} \mathbf{u} }{ \mathrm{d} \psi }(T),
\end{equation*}
contracted against a vector, $\delta$,
\begin{equation*}
\boldsymbol{\delta}^{\dagger} \cdot 
\frac{ \mathrm{d} \mathbf{u} }{ \mathrm{d} \psi}(T),
\end{equation*}
where $\dagger$ denotes transposition.  This arises, for example, when computing the 
gradient of a scalar function, for example a probability density or an objective function,
which implicitly depends on $\psi$ through $\mathbf u$.

We can recover the above contraction by defining the integrand
\begin{equation*}
j(\mathbf{u}, \psi, t) = \boldsymbol{\delta}^{\dagger} \cdot \mathbf{f} (\mathbf{u}, \psi, t)
\end{equation*}
and the corresponding functional
\begin{align*}
\mathcal{J}(\psi) 
&=
\int_{0}^{T} \mathrm{d} t \, j(\mathbf{u}, \psi, t)
\\
&=
\boldsymbol{\delta}^{\dagger} \cdot \int_{0}^{T} \mathrm{d} t \, \mathbf{f} (\mathbf{u}, \psi, t)
\\
&=
\boldsymbol{\delta}^{\dagger} \cdot \int_{0}^{T} \mathrm{d} t \, 
\frac{ \mathrm{d} \mathbf{u} }{ \mathrm{d} t } (\mathbf{u}, \psi, t)
\\
&=
\boldsymbol{\delta}^{\dagger} \cdot \left( \mathbf{u}(T) - \mathbf{u}(0) \right).
\end{align*}
The total derivative of this functional is given by
\begin{align*}
\frac{ \mathrm{d} \mathcal{J} }{ \mathrm{d} \psi}(\psi)
&=
\boldsymbol{\delta}^{\dagger} \cdot \left( 
\frac{ \mathrm{d} \mathbf{u} }{ \mathrm{d} \psi} (T)
- \frac{ \mathrm{d} \mathbf{u} }{ \mathrm{d} \psi} (0) \right)
\\
&=
\boldsymbol{\delta}^{\dagger} \cdot \left( 
\frac{ \mathrm{d} \mathbf{u} }{ \mathrm{d} \psi} (T)
- \frac{ \partial \boldsymbol{\upsilon} }{ \partial \psi} \right)
\end{align*}
which we can then manipulate into the desired contraction
\begin{equation*}
\boldsymbol{\delta}^{\dagger} \cdot
\frac{ \mathrm{d} \mathbf{u} }{ \mathrm{d} \psi} (T)
=
\frac{ \mathrm{d} \mathcal{J} }{ \mathrm{d} \psi}(\psi)
+ \boldsymbol{\delta}^{\dagger} \cdot
\frac{ \partial \boldsymbol{\upsilon} }{ \partial \psi}.
\end{equation*}

We can then use the continuous adjoint method to evaluate the total 
derivative of the functional and hence the desired Jacobian-adjoint 
product,
\begin{align*}
\boldsymbol{\delta}^{\dagger} \cdot
\frac{ \mathrm{d} \mathbf{u} }{ \mathrm{d} \psi} (T)
&=
\boldsymbol{\delta}^{\dagger} \cdot
\frac{ \partial \boldsymbol{\upsilon} }{ \partial \psi}
+ \frac{ \mathrm{d} \mathcal{J} }{ \mathrm{d} \psi}(\psi)
\\
&=
\boldsymbol{\delta}^{\dagger} \cdot
\frac{ \partial \boldsymbol{\upsilon} }{ \partial \psi}
- \boldsymbol{\lambda}^{\dagger}(0) \cdot \frac{ \partial \boldsymbol{\upsilon}}{ \partial \psi}
+ \int_{0}^{T} \mathrm{d}t \, 
\boldsymbol{\delta}^{\dagger} \cdot \frac{ \partial \mathbf{f} }{ \partial \psi}(\mathbf{u}, \psi, t)
- \boldsymbol{\lambda}^{\dagger}(t) \cdot
\frac{ \partial \mathbf{f} }{ \partial \mathrm{\psi} }(\mathbf{u}, \psi, t)
\\
&=
\bigg[ \boldsymbol{\delta} - \boldsymbol{\lambda}(0) \bigg]^{\dagger}
\cdot \frac{ \partial \boldsymbol{\upsilon}}{ \partial \psi}
+ \int_{0}^{T} \mathrm{d}t \, 
\bigg[ \boldsymbol{\delta} - \boldsymbol{\lambda}(t) \bigg]^{\dagger} \cdot
\frac{ \partial \mathbf{f} }{ \partial \mathrm{\psi} }(\mathbf{u}, \psi, t).
\end{align*}

\section{Discrete Adjoint Systems}

By carefully translating the differential operations in the continuous adjoint method 
to their discrete counterparts we can derive a corresponding discrete adjoint method.

Recall that in the discrete case our target functional is defined as
\begin{equation*}
\mathcal{J}(\psi) = \sum_{n = 0}^{N - 1} j_{n}(\mathbf{u}_{n}, \psi, n)
\end{equation*}
with the discrete states satisfying the forward difference equation,
\begin{equation*}
\mathbf{u}_{n + 1} - \mathbf{u}_{n} = \boldsymbol{\Delta}_{n} (\mathbf{u}_{n}, \psi, n),
\end{equation*}
along with the initial condition
\begin{equation*}
\mathbf{u}_{0}(\psi) = \boldsymbol{\upsilon}(\psi).
\end{equation*}

To construct the adjoint system we first introduce the nominal system as explicit
constraints in a modified functional,
\begin{equation*}
\mathcal{J}(\psi) = \mathcal{L}(\psi) = 
\boldsymbol{\mu}^{T} \cdot \left[ \boldsymbol{\upsilon} - \mathbf{u_0} \right]
+ \sum_{n = 0}^{N - 1} 
j_{n}
+ \boldsymbol{\lambda}_{n}^{T} \cdot 
\left[ \mathbf{u}_{n + 1} - \mathbf{u}_{n} - \boldsymbol{\Delta}_{n} \right].
\end{equation*}
Taking a total derivative then gives
\begin{align*}
\frac{ \mathrm{d} \mathcal{J} }{ \mathrm{d} \psi}
&=
\frac{ \mathrm{d} \mathcal{L} }{ \mathrm{d} \psi}
\\
&= \quad
\boldsymbol{\mu}^{\dagger} \cdot \left[ 
\frac{ \mathrm{d} \mathbf{u}_{0} }{ \mathrm{d} \psi}
- \frac{ \mathrm{d} \boldsymbol{\upsilon} }{ \mathrm{d} \psi} \right]
+ \sum_{n = 0}^{N - 1} 
\frac{ \mathrm{d} j_{n} }{ \mathrm{d} \psi}
+ \boldsymbol{\lambda}_{n}^{\dagger} \cdot 
\left[ \frac{ \mathrm{d} \mathbf{u}_{n + 1} }{ \mathrm{d} \psi} 
- \frac{ \mathrm{d} \mathbf{u}_{n} }{\mathrm{d} \psi } 
- \frac{ \mathrm{d} \boldsymbol{\Delta}_{n} }{ \mathrm{d} \psi} \right]
\\
&= \quad
\boldsymbol{\mu}^{\dagger} \cdot \left[ 
\frac{ \mathrm{d} \mathbf{u}_{0} }{ \mathrm{d} \psi}
- \frac{ \mathrm{d} \boldsymbol{\upsilon} }{ \mathrm{d} \psi} \right]
\\
&\quad + \sum_{n = 0}^{N - 1} 
\frac{ \partial j_{n} }{ \partial \psi}
+
\left( \frac{ \partial j_{n} }{ \partial \mathbf{u}_{n} } \right)^{\dagger}
\cdot
\frac{ \mathrm{d} \mathbf{u}_{n} }{ \mathrm{d} \psi }
+ \boldsymbol{\lambda}_{n}^{\dagger} \cdot 
\left[ \frac{ \mathrm{d} \mathbf{u}_{n + 1} }{ \mathrm{d} \psi} 
- \frac{ \mathrm{d} \mathbf{u}_{n} }{\mathrm{d} \psi } 
- \frac{ \partial \boldsymbol{\Delta}_{n} }{ \partial \psi}
- \left( \frac{ \partial \boldsymbol{\Delta}_{n} }{ \partial \mathbf{u}_{n} } \right)^{\dagger}
\cdot \frac{ \mathrm{d} \mathbf{u}_{n} }{ \mathrm{d} \psi}
\right]
\\
&= \quad
\boldsymbol{\mu}^{\dagger} \cdot \left[ 
\frac{ \mathrm{d} \mathbf{u}_{0} }{ \mathrm{d} \psi}
- \frac{ \mathrm{d} \boldsymbol{\upsilon} }{ \mathrm{d} \psi} \right]
\\
&\quad 
+ \sum_{n = 0}^{N - 1} 
\frac{ \partial j_{n} }{ \partial \psi}
- \boldsymbol{\lambda}_{n}^{\dagger} \cdot \frac{ \partial \boldsymbol{\Delta}_{n} }{ \partial \psi}
+ \sum_{n = 0}^{N - 1} 
\boldsymbol{\lambda}_{n}^{\dagger} \cdot 
\left[ \frac{ \mathrm{d} \mathbf{u}_{n + 1} }{ \mathrm{d} \psi} 
- \frac{ \mathrm{d} \mathbf{u}_{n} }{\mathrm{d} \psi } 
\right]
+ \sum_{n = 0}^{N - 1} 
\left( \frac{ \partial j_{n} }{ \partial \mathbf{u}_{n} } \right)^{\dagger}
\cdot
\frac{ \mathrm{d} \mathbf{u}_{n} }{ \mathrm{d} \psi }
- \left( \boldsymbol{\lambda}_{n} \cdot 
\frac{ \partial \boldsymbol{\Delta}_{n} }{ \partial \mathbf{u}_{n} } \right)^{\dagger}
\cdot \frac{ \mathrm{d} \mathbf{u}_{n} }{ \mathrm{d} \psi}
\\
&= \quad
\boldsymbol{\mu}^{\dagger} \cdot \left[ 
\frac{ \mathrm{d} \mathbf{u}_{0} }{ \mathrm{d} \psi}
- \frac{ \mathrm{d} \boldsymbol{\upsilon} }{ \mathrm{d} \psi} \right]
+ \left( \frac{ \partial j_{0} }{ \partial \mathbf{u}_{0} } \right)^{\dagger}
\cdot \frac{ \mathrm{d} \mathbf{u}_{0} }{ \mathrm{d} \psi }
- \left( \boldsymbol{\lambda}_{0} \cdot 
\frac{ \partial \boldsymbol{\Delta}_{0} }{ \partial \mathbf{u}_{0} } \right)^{\dagger}
\cdot \frac{ \mathrm{d} \mathbf{u}_{0} }{ \mathrm{d} \psi}
\\
&\quad 
+ \sum_{n = 0}^{N - 1}
\frac{ \partial j_{n} }{ \partial \psi}
- \boldsymbol{\lambda}_{n}^{\dagger} \cdot \frac{ \partial \boldsymbol{\Delta}_{n} }{ \partial \psi}
+ \sum_{n = 0}^{N - 1} 
\boldsymbol{\lambda}_{n}^{\dagger} \cdot 
\left[ \frac{ \mathrm{d} \mathbf{u}_{n + 1} }{ \mathrm{d} \psi} 
- \frac{ \mathrm{d} \mathbf{u}_{n} }{\mathrm{d} \psi } 
\right]
+ \sum_{n = 1}^{N - 1} 
\left( \frac{ \partial j_{n} }{ \partial \mathbf{u}_{n} } \right)^{\dagger}
\cdot
\frac{ \mathrm{d} \mathbf{u}_{n} }{ \mathrm{d} \psi }
- \left( \boldsymbol{\lambda}_{n} \cdot 
\frac{ \partial \boldsymbol{\Delta}_{n} }{ \partial \mathbf{u}_{n} } \right)^{\dagger}
\cdot \frac{ \mathrm{d} \mathbf{u}_{n} }{ \mathrm{d} \psi}.
\end{align*}

Now we can apply \emph{summation by parts} to the forwards difference of
sensitivities,
\begin{align*}
\sum_{n = 0}^{N - 1} 
\boldsymbol{\lambda}_{n}^{\dagger} \cdot 
\left[ \frac{ \mathrm{d} \mathbf{u}_{n + 1} }{ \mathrm{d} \psi} 
- \frac{ \mathrm{d} \mathbf{u}_{n} }{\mathrm{d} \psi } 
\right]
&=
\boldsymbol{\lambda}_{N - 1}^{\dagger} \cdot \frac{ \mathrm{d} \mathbf{u}_{N} }{ \mathrm{d} \psi} 
- \boldsymbol{\lambda}_{0}^{\dagger} \cdot \frac{ \mathrm{d} \mathbf{u}_{0} }{ \mathrm{d} \psi} 
- \sum_{n = 1}^{N - 1} 
\bigg[ \boldsymbol{\lambda}_{n} - \boldsymbol{\lambda}_{n - 1} \bigg]^{\dagger}
\cdot \frac{ \mathrm{d} \mathbf{u}_{n} }{ \mathrm{d} \psi }.
\end{align*}
Plugging this result into our functional derivative then gives
\begin{align*}
\frac{ \mathrm{d} \mathcal{J} }{ \mathrm{d} \psi}
&= \quad
\boldsymbol{\mu}^{\dagger} \cdot \left[ 
\frac{ \mathrm{d} \mathbf{u}_{0} }{ \mathrm{d} \psi}
- \frac{ \mathrm{d} \boldsymbol{\upsilon} }{ \mathrm{d} \psi} \right]
+ \left( \frac{ \partial j_{0} }{ \partial \mathbf{u}_{0} } \right)^{\dagger}
\cdot \frac{ \mathrm{d} \mathbf{u}_{0} }{ \mathrm{d} \psi }
- \left( \boldsymbol{\lambda}_{0} \cdot 
\frac{ \partial \boldsymbol{\Delta}_{0} }{ \partial \mathbf{u}_{0} } \right)^{\dagger}
\cdot \frac{ \mathrm{d} \mathbf{u}_{0} }{ \mathrm{d} \psi}
+ \boldsymbol{\lambda}_{N - 1}^{\dagger} \cdot \frac{ \mathrm{d} \mathbf{u}_{N} }{ \mathrm{d} \psi } 
- \boldsymbol{\lambda}_{0}^{\dagger} \cdot \frac{ \mathrm{d} \mathbf{u}_{0} }{ \mathrm{d} \psi} 
\\
&\quad + \sum_{n = 0}^{N - 1} 
\frac{ \partial j_{n} }{ \partial \psi}
- \boldsymbol{\lambda}_{n}^{\dagger} \cdot 
\frac{ \partial \boldsymbol{\Delta}_{n} }{ \partial \psi}
\\
&\quad + \sum_{n = 1}^{N - 1} 
\left( \frac{ \partial j_{n} }{ \partial \mathbf{u}_{n} } \right)^{\dagger}
\cdot
\frac{ \mathrm{d} \mathbf{u}_{n} }{ \mathrm{d} \psi }
- \left( \boldsymbol{\lambda}_{n} \cdot
\frac{ \partial \boldsymbol{\Delta}_{n} }{ \partial \mathbf{u}_{n} } \right)^{\dagger}
\cdot \frac{ \mathrm{d} \mathbf{u}_{n} }{ \mathrm{d} \psi}
- \sum_{n = 1}^{N - 1} 
\bigg[ \boldsymbol{\lambda}_{n} - \boldsymbol{\lambda}_{n - 1} \bigg]^{\dagger}
\cdot \frac{ \mathrm{d} \mathbf{u}_{n} }{ \mathrm{d} \psi }
\\
&= \quad
\left[
\boldsymbol{\mu} + \frac{ \partial j_{0} }{ \partial \boldsymbol{\upsilon} }
- \boldsymbol{\lambda}_{0} \cdot \frac{ \partial \boldsymbol{\Delta}_{0} }{ \partial \boldsymbol{\upsilon} }
- \boldsymbol{\lambda}_{0}
\right]^{\dagger} 
\cdot \frac{ \mathrm{d} \mathbf{u}_{0} }{ \mathrm{d} \psi}
- \boldsymbol{\mu}^{\dagger} \cdot \frac{ \partial \boldsymbol{\upsilon} }{ \partial \psi }
+ \boldsymbol{\lambda}_{N - 1}^{\dagger} \cdot \frac{ \mathrm{d} \mathbf{u}_{N} }{ \mathrm{d} \psi } 
\\
&\quad + \sum_{n = 0}^{N - 1} 
\frac{ \partial j_{n} }{ \partial \psi}
- \boldsymbol{\lambda}_{n}^{\dagger} \cdot 
\frac{ \partial \boldsymbol{\Delta}_{n} }{ \partial \psi}
\\
&\quad + \sum_{n = 1}^{N - 1} 
\left[ 
\frac{ \partial j_{n} }{ \partial \mathbf{u}_{n} }
- 
\boldsymbol{\lambda}_{n} + \boldsymbol{\lambda}_{n - 1}
- \boldsymbol{\lambda}_{n} \cdot
\frac{ \partial \boldsymbol{\Delta}_{n} }{ \partial \mathbf{u}_{n} }
\right]^{\dagger} \cdot \frac{ \mathrm{d} \mathbf{u}_{n} }{ \mathrm{d} \psi }.
\end{align*}

As in the discrete case we can exploit the freedom in our Lagrange multipliers to remove 
all of the sensitivity terms.  We first set
\begin{equation*}
\boldsymbol{\mu} + \frac{ \partial j_{0} }{ \partial \mathbf{u}_{0} }
- \boldsymbol{\lambda}_{0} \cdot \frac{ \partial \boldsymbol{\Delta}_{0} }{ \partial \mathbf{u}_{0} }
- \boldsymbol{\lambda}_{0}
= 0,
\end{equation*}
or
\begin{equation*}
\boldsymbol{\mu} =
- \frac{ \partial j_{0} }{ \partial \mathbf{u}_{0} }
+ \boldsymbol{\lambda}_{0} \cdot \frac{ \partial \boldsymbol{\Delta}_{0} }{ \partial \mathbf{u}_{0} }
+ \boldsymbol{\lambda}_{0},
\end{equation*}
and then 
\begin{equation*}
\boldsymbol{\lambda}_{N - 1} = 0
\end{equation*}
to remove all the sensitivities outside of the summations.  We then eliminate the second summation 
by choosing the rest of the $\boldsymbol{\lambda}_{n}$ to satisfy 
\begin{equation*}
\frac{ \partial j_{n} }{ \partial \mathbf{u}_{n} }
- 
\boldsymbol{\lambda}_{n} + \boldsymbol{\lambda}_{n - 1}
- \boldsymbol{\lambda}_{n} \cdot
\frac{ \partial \boldsymbol{\Delta}_{n} }{ \partial \mathbf{u}_{n} }
= 0,
\end{equation*}
or equivalently
\begin{equation*}
\frac{ \partial j_{n + 1} }{ \partial \mathbf{u}_{n + 1} }
- 
\boldsymbol{\lambda}_{n + 1} + \boldsymbol{\lambda}_{n}
- \boldsymbol{\lambda}_{n + 1} \cdot
\frac{ \partial \boldsymbol{\Delta}_{n + 1} }{ \partial \mathbf{u}_{n + 1} }
= 0.
\end{equation*}

This defines an adjoint system defined by the \emph{backward} difference equations
\begin{equation*}
\boldsymbol{\lambda}_{n} - \boldsymbol{\lambda}_{n + 1}
=
- \frac{ \partial j_{n + 1} }{ \partial \mathbf{u}_{n + 1} }
+ \boldsymbol{\lambda}_{n + 1} \cdot 
\frac{ \partial \boldsymbol{\Delta}_{n + 1} }{ \partial \mathbf{u}_{n + 1} }
\end{equation*}
along with the terminal condition 
\begin{equation*}
\boldsymbol{\lambda}_{N - 1} = 0.
\end{equation*}
If we solve for the sequence $\boldsymbol{\lambda}_{(N-1):0}$ after first forward solving
the original sequence $\mathbf{u}_{0:N}$, we can compute the total derivative of the 
functional as
\begin{equation*}
\frac{ \mathrm{d} \mathcal{J} }{ \mathrm{d} \psi}
=
\left[ \frac{ \partial j_{0} }{ \partial \mathbf{u}_{0} }
- \boldsymbol{\lambda}_{0} \cdot \frac{ \partial \boldsymbol{\Delta}_{0} }{ \partial \mathbf{u}_{0} }
- \boldsymbol{\lambda}_{0} \right]^{\dagger} \cdot \frac{ \partial \mathbf{\upsilon} }{ \partial \psi }
+ \sum_{n = 0}^{N - 1} 
\frac{ \partial j_{n} }{ \partial \psi}
-\boldsymbol{\lambda}_{n}^{\dagger} \cdot 
\frac{ \partial \boldsymbol{\Delta}_{n} }{ \partial \psi}.
\end{equation*}

\section{Application to Hidden Markov Models}

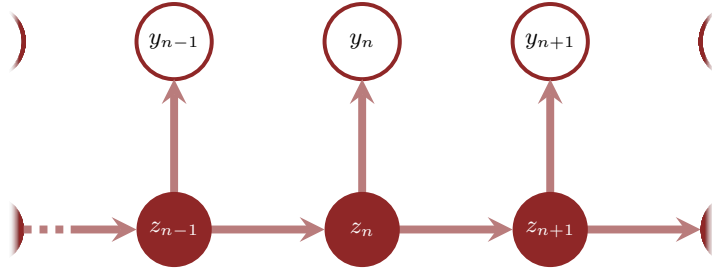
\begin{figure}
\centering
\begin{tikzpicture}[scale=0.5, thick]

\foreach \i in {0, 0.01,..., 1} {
  \begin{scope}
    \clip (6.075, -1) rectangle +(-0.5 * \i, 2);
    \fill[color=dark, line width=1.5, opacity={exp(-6 * \i)}] (5, 0) circle (1);
  \end{scope}
  
  \begin{scope}
    \clip (6.075, 4) rectangle +(-0.5 * \i, 2);
    \draw[color=dark, line width=1.5, opacity={exp(-6 * \i)}] (5, 5) circle (1);
  \end{scope}
}  

\draw[-, color=mid, line width=3, dashed] (6, 0) -- +(1.25, 0);
\draw[->, >=stealth, color=mid, line width=3] (7.25, 0) -- +(1.75, 0);

\filldraw[fill=white, draw=dark, line width=1.5] (10, 5) circle (1)
node[color=black] { $y_{n - 1}$ };

\draw[->, >=stealth, color=mid, line width=3] (10, 0) -- +(0, 4);
\draw[->, >=stealth, color=mid, line width=3] (10, 0) -- +(4, 0);

\fill[color=dark] (10, 0) circle (1)
node[color=white] { $z_{n - 1}$ };

\filldraw[fill=white, draw=dark, line width=1.5] (15, 5) circle (1)
node[color=black] { $y_{n}$ };

\draw[->, >=stealth, color=mid, line width=3] (15, 0) -- +(0, 4);
\draw[->, >=stealth, color=mid, line width=3] (15, 0) -- +(4, 0);

\fill[color=dark] (15, 0) circle (1)
node[color=white] { $z_{n}$ };

\filldraw[fill=white, draw=dark, line width=1.5] (20, 5) circle (1)
node[color=black] { $y_{n + 1}$ };

\draw[->, >=stealth, color=mid, line width=3] (20, 0) -- +(0, 4);
\draw[->, >=stealth, color=mid, line width=3] (20, 0) -- +(4, 0);

\fill[color=dark] (20, 0) circle (1)
node[color=white] { $z_{n + 1}$ };

\foreach \i in {0, 0.01,..., 1} {
  \begin{scope}
    \clip (23.925, -1) rectangle +(0.5 * \i, 2);
    \fill[color=dark, line width=1.5, opacity={exp(-6 * \i)}] (25, 0) circle (1);
  \end{scope}
  
  \begin{scope}
    \clip (23.925, 4) rectangle +(0.5 * \i, 2);
    \draw[color=dark, line width=1.5, opacity={exp(-6 * \i)}] (25, 5) circle (1);
  \end{scope}
}

\end{tikzpicture}
\caption{The conditional dependence structure of a hidden Markov model admits 
efficient marginalization of the discrete hidden states into state probabilities.  
Derivatives of the state probabilities with respect to the model parameters also 
have to navigate this conditional dependence structure.}
\label{fig:hmmc}
\end{figure}

The discrete adjoint method is applicable to any discrete sequence defined by 
forward difference equations that depend only on the current state.  In this
section we demonstrate an application of the method to common hidden Markov 
models.

An elementary hidden Markov model is a probabilistic model over $N$ observations, $y_{n}$, 
and $N$ hidden states, $z_{n}$, satisfying the conditional dependence structure shown 
in Figure~\ref{fig:hmmc}.  The joint density $\pi(y_{1:N}, z_{1:N}, \psi)$ is readily 
computed, but the derivatives are ill-defined when the hidden states $z$ are discrete.  
In order to apply gradient-based methods we first need to marginalize out the hidden 
states to define the \emph{marginal likelihood} $\pi(y_{1:N}, \psi)$ which can be 
differentiated.

Fortunately exact marginalization is tractable due to the conditional dependencies 
inherent to a hidden Markov model.  Defining the observational density functions
\begin{equation*}
\omega_{n, i} \equiv \pi(y_n \mid z_n = i)
\end{equation*}
and the transition matrices
\begin{equation*}
\Gamma_{n, ij} \equiv \pi(z_{n + 1} = i \mid z_n = j)
\end{equation*}
we can marginalize the hidden states into the forward state probabilities 
\begin{equation*}
\alpha_{n, i} \equiv \pi(y_{1:N}, z_n = i).
\end{equation*}
Because of the defining conditional structure these state probabilities satisfy 
the recursion relation
\begin{equation*}
\boldsymbol{\alpha}_{n + 1}(\psi) = 
\boldsymbol{\omega}_{n + 1}(\psi)
\circ ( \boldsymbol{\Gamma}_{n + 1}(\psi) \cdot \boldsymbol{\alpha}_{n}(\psi)),
\end{equation*}
where $\circ$ denotes the element-wise Hadamard product, along with the 
initial condition
\begin{equation*}
\boldsymbol{\upsilon}(\psi)
= \boldsymbol{\alpha}_{0}(\psi)
= \boldsymbol{\omega}_{0}(\psi) \circ \boldsymbol{\rho}(\psi).
\end{equation*}
Forward solving the recursion relation efficiently computes each of the 
state probabilities, the last of which gives the desired marginal likelihood
\begin{equation*}
\pi(y_{1}, \ldots, y_{N}, \psi)
=
\sum_{m = 1}^{M} \alpha_{N, m} (\psi)
=
\mathbf{1}^{\dagger} \cdot \boldsymbol{\alpha}_{N} (\psi).
\end{equation*}
In order to apply gradient-based learning algorithms to any probabilistic 
model containing a hidden Markov model we have to compute not only the marginal 
likelihood but also its gradient with respect to any unknown parameters.

There are many ways to derive the gradient for this problem; in this section we will consider 
three approaches that tackle the derivation from different directions and different intuitions
but arrive at the same result.  These different approaches not only serve as cross checks for 
each other but also suggest that their common result is optimal.

In the statistics literature the gradient of the marginal likelihood is often derived as an 
indirect and subtle byproduct of the expectation maximization algorithm \citep{CappeEtAl:2005}. 

We can also obtain a more explicit derivation by unrolling the recursion and applying 
the chain rule iteratively.  If we let $\boldsymbol{\Omega}_{n}$ denote a 
diagonal matrix of observational densities at the $n$th iteration,
\begin{equation*}
\boldsymbol{\Omega}_{n} = \mathrm{diag}( \boldsymbol{\omega}_{n} ),
\end{equation*}
then the final state probabilities can be written explicitly as
\begin{equation*}
\boldsymbol{\alpha}_{N}
=
\left[ \prod_{n = 1}^{N} \boldsymbol{\Omega}_{n}(\psi) \cdot \boldsymbol{\Gamma}_{n + 1}(\psi) \right]
\cdot \boldsymbol{\Omega}_{0}(\psi) \cdot \boldsymbol{\rho}(\psi)
\end{equation*}
with the marginal likelihood taking the form
\begin{equation*}
\pi(y_{1}, \ldots, y_{N}, \psi) 
=
\mathbf{1}^{\dagger} \cdot \boldsymbol{\alpha}_{N}
=
\mathbf{1}^{\dagger} \cdot
\left[ \prod_{n = 1}^{N} \boldsymbol{\Omega}_{n} \cdot \boldsymbol{\Gamma}_{n + 1} \right]
\cdot \boldsymbol{\Omega}_{0} \cdot \boldsymbol{\rho}.
\end{equation*}
Applying the product rule for derivatives then gives
\begin{align*}
\frac{ \mathrm{d} }{ \mathrm{d} \psi }
\pi(y_{1}, \ldots, y_{N}) 
&=
\frac{ \mathrm{d} }{ \mathrm{d} \psi } \left(
\mathbf{1}^{\dagger} \cdot
\left[ \prod_{n = 1}^{N - 1} \boldsymbol{\Omega}_{n} \cdot \boldsymbol{\Gamma}_{n + 1} \right]
\cdot \boldsymbol{\Omega}_{0} \cdot \boldsymbol{\rho} \right)
\\
&= \quad
\mathbf{1}^{\dagger} \cdot
\sum_{j = 0}^{N - 1}
\left[ \prod_{i = j + 2}^{N - 1} \boldsymbol{\Omega}_{i} \cdot \boldsymbol{\Gamma}_{i + 1} \right]
\cdot
\Bigg[
\frac{ \mathrm{d} \boldsymbol{\Omega}_{j + 1} }{ \mathrm{d} \psi } \cdot \boldsymbol{\Gamma}_{j + 2}
+
\boldsymbol{\Omega}_{j + 1} \cdot \frac{ \mathrm{d} \boldsymbol{\Gamma}_{j + 2} }{ \mathrm{d} \psi } 
\Bigg]
\cdot
\left[ \prod_{k = 1}^{j} \boldsymbol{\Omega}_{k} \cdot \boldsymbol{\Gamma}_{k + 1} \right]
\cdot \boldsymbol{\Omega}_{0} \cdot \boldsymbol{\rho}
\\
& \quad +
\mathbf{1}^{\dagger} \cdot
\left[ \prod_{n = 1}^{N - 1} \boldsymbol{\Omega}_{n} \cdot \boldsymbol{\Gamma}_{n + 1} \right] \cdot
\Bigg[ \frac{ \mathrm{d} \boldsymbol{\Omega}_{0} }{ \mathrm{d} \psi} \cdot \boldsymbol{\rho}
+
\boldsymbol{\Omega}_{0} \cdot \frac{ \mathrm{d} \boldsymbol{\rho} }{ \mathrm{d} \psi} \Bigg]
\\
&= \quad
\sum_{j = 0}^{N - 1}
\left[ 
\mathbf{1}^{\dagger} \cdot
\prod_{i = j + 2}^{N - 1} \boldsymbol{\Omega}_{i} \cdot \boldsymbol{\Gamma}_{i + 1} \right]
\cdot
\Bigg[
\frac{ \mathrm{d} \boldsymbol{\Omega}_{j + 1} }{ \mathrm{d} \psi } \cdot \boldsymbol{\Gamma}_{j + 2}
+
\boldsymbol{\Omega}_{j + 1} \cdot \frac{ \mathrm{d} \boldsymbol{\Gamma}_{j + 2} }{ \mathrm{d} \psi } 
\Bigg]
\cdot
\boldsymbol{\alpha}_{j}
\\
& \quad +
\mathbf{1}^{\dagger} \cdot
\left[ \prod_{n = 1}^{N - 1} \boldsymbol{\Omega}_{n} \cdot \boldsymbol{\Gamma}_{n + 1} \right] \cdot
\Bigg[ \frac{ \mathrm{d} \boldsymbol{\Omega}_{0} }{ \mathrm{d} \psi} \cdot \boldsymbol{\rho}
+
\boldsymbol{\Omega}_{0} \cdot \frac{ \mathrm{d} \boldsymbol{\rho} }{ \mathrm{d} \psi} \Bigg]
\\
&= \quad
\sum_{j = 0}^{N - 1}
\left[ \left[ 
\mathbf{1}^{\dagger} \cdot
\prod_{i = j + 2}^{N - 1} \boldsymbol{\Omega}_{i} \cdot \boldsymbol{\Gamma} _{i + 1}
\right]^{\dagger} \right]^{\dagger}
\cdot
\Bigg[
\frac{ \mathrm{d} \boldsymbol{\Omega}_{j + 1} }{ \mathrm{d} \psi } \cdot \boldsymbol{\Gamma}_{j + 2}
+
\boldsymbol{\Omega}_{j + 1} \cdot \frac{ \mathrm{d} \boldsymbol{\Gamma}_{j + 2} }{ \mathrm{d} \psi } 
\Bigg]
\cdot
\boldsymbol{\alpha}_{j}
\\
& \quad +
\left[ \left[ 
\mathbf{1}^{\dagger} \cdot
\prod_{i = 1}^{N - 1} \boldsymbol{\Omega}_{i} \cdot \boldsymbol{\Gamma}_{i + 1}
\right]^{\dagger} \right]^{\dagger} \cdot
\Bigg[ \frac{ \mathrm{d} \boldsymbol{\Omega}_{0} }{ \mathrm{d} \psi} \cdot \boldsymbol{\rho}
+
\boldsymbol{\Omega}_{0} \cdot \frac{ \mathrm{d} \boldsymbol{\rho} }{ \mathrm{d} \psi} \Bigg]
\\
&=
\sum_{j = 0}^{N - 1}
\left[ \left[ 
\prod_{i = N - 1}^{j + 2} \boldsymbol{\Gamma}^{\dagger}_{i + 1} \cdot \boldsymbol{\Omega}_{i}^{\dagger} \right]
\cdot \mathbf{1}
 \right]^{\dagger}
\cdot
\Bigg[
\frac{ \mathrm{d} \boldsymbol{\Omega}_{j + 1} }{ \mathrm{d} \psi } \cdot \boldsymbol{\Gamma}_{j + 2}
+
\boldsymbol{\Omega}_{j + 1} \cdot \frac{ \mathrm{d} \boldsymbol{\Gamma}_{j + 2} }{ \mathrm{d} \psi } 
\Bigg]
\cdot
\boldsymbol{\alpha}_{j}
\\
& \quad +
\left[ \left[ 
\prod_{i = N - 1}^{1} \boldsymbol{\Gamma}^{\dagger}_{i + 1} \cdot \boldsymbol{\Omega}_{i}^{\dagger} \right]
\cdot \mathbf{1}
\right]^{\dagger} \cdot
\Bigg[ \frac{ \mathrm{d} \boldsymbol{\Omega}_{0} }{ \mathrm{d} \psi} \cdot \boldsymbol{\rho}
+
\boldsymbol{\Omega}_{0} \cdot \frac{ \mathrm{d} \boldsymbol{\rho} }{ \mathrm{d} \psi} \Bigg]
\\
&=
\sum_{j = 0}^{N - 1}
\left[ \left[ 
\prod_{i = N - 1}^{j + 2} \boldsymbol{\Gamma}^{\dagger}_{i + 1} \cdot \boldsymbol{\Omega}_{i} \right]
\cdot \mathbf{1}
 \right]^{\dagger}
\cdot
\Bigg[
\frac{ \mathrm{d} \boldsymbol{\Omega}_{j + 1} }{ \mathrm{d} \psi } \cdot \boldsymbol{\Gamma}_{j + 2}
+
\boldsymbol{\Omega}_{j + 1} \cdot \frac{ \mathrm{d} \boldsymbol{\Gamma}_{j + 2} }{ \mathrm{d} \psi } 
\Bigg]
\cdot
\boldsymbol{\alpha}_{j}
\\
& \quad +
\left[ \left[ 
\prod_{i = N - 1}^{1} \boldsymbol{\Gamma}^{\dagger}_{i + 1} \cdot \boldsymbol{\Omega}_{i} \right]
\cdot \mathbf{1}
\right]^{\dagger} \cdot
\Bigg[ \frac{ \mathrm{d} \boldsymbol{\Omega}_{0} }{ \mathrm{d} \psi} \cdot \boldsymbol{\rho}
+
\boldsymbol{\Omega}_{0} \cdot \frac{ \mathrm{d} \boldsymbol{\rho} }{ \mathrm{d} \psi} \Bigg]
\\
&=
\sum_{j = 0}^{N - 1}
\Bigg[ \boldsymbol{\beta}_{j + 1} \Bigg]^{\dagger}
\cdot
\Bigg[
\frac{ \mathrm{d} \boldsymbol{\Omega}_{j + 1} }{ \mathrm{d} \psi } \cdot \boldsymbol{\Gamma}_{j + 2}
+
\boldsymbol{\Omega}_{j + 1} \cdot \frac{ \mathrm{d} \boldsymbol{\Gamma}_{j + 2} }{ \mathrm{d} \psi } 
\Bigg]
\cdot
\boldsymbol{\alpha}_{j}
\\
& \quad +
\Bigg[ \boldsymbol{\beta}_{0} \Bigg]^{\dagger} \cdot
\Bigg[ \frac{ \mathrm{d} \boldsymbol{\Omega}_{0} }{ \mathrm{d} \psi} \cdot \boldsymbol{\rho}
+
\boldsymbol{\Omega}_{0} \cdot \frac{ \mathrm{d} \boldsymbol{\rho} }{ \mathrm{d} \psi} \Bigg],
\end{align*}
where we have defined the \emph{backwards states}
\begin{equation*}
\boldsymbol{\beta}_{j} 
=
\left[ 
\prod_{i = N - 1}^{j + 1} \boldsymbol{\Gamma}^{\dagger}_{i + 1} \cdot \boldsymbol{\Omega}_{i} \right]
\cdot \mathbf{1}
\end{equation*}

A third, novel approach to deriving the marginal likelihood gradient is to interpret the recursion 
as a forward difference equation and apply the discrete adjoint method.  Let 
$\mathbf{u}_{n} = \boldsymbol{\alpha}_{n}$ and manipulate the defining recursion relation into a 
forward difference
\begin{equation*}
\boldsymbol{\Delta}_{n}
=
\boldsymbol{\omega}_{n + 1}
\circ ( \boldsymbol{\Gamma}_{n + 1} \cdot \boldsymbol{\alpha}_{n} ) - \boldsymbol{\alpha}_{n},
\end{equation*}
and take the summand
\begin{equation*}
j = \mathbf{1}^{\dagger} \cdot \boldsymbol{\Delta}_{n}
\end{equation*}
to give the discrete functional
\begin{equation*}
J = \mathbf{1}^{\dagger} \cdot \left( \boldsymbol{\alpha}_{N} - \boldsymbol{\upsilon} \right).
\end{equation*}

The total derivative of the discrete functional can be used to derive the derivative of the 
marginal likelihood,
\begin{align*}
\frac{ \mathrm{d} \mathcal{J} }{ \mathrm{d} \psi}
&=
\frac{\mathrm{d}}{ \mathrm{d} \psi} \left( \mathbf{1}^{\dagger} \cdot \boldsymbol{\alpha}_{N} \right)
- \frac{\mathrm{d}}{ \mathrm{d} \psi} \left( \mathbf{1}^{\dagger} \cdot \boldsymbol{\alpha}_{0} \right)
\\
&=
\frac{\mathrm{d}}{ \mathrm{d} \psi} \pi(y_{1}, \ldots, y_{N})
- \mathbf{1}^{\dagger} \cdot \frac{\mathrm{d} \boldsymbol{\alpha}_{0} }{ \mathrm{d} \psi},
\end{align*}
or
\begin{equation*}
\frac{\mathrm{d}}{ \mathrm{d} \psi} \pi(y_{1}, \ldots, y_{N})
= \frac{ \mathrm{d} \mathcal{J} }{ \mathrm{d} \psi}
+ \mathbf{1}^{\dagger} \cdot \frac{\mathrm{d} \boldsymbol{\alpha}_{0} }{ \mathrm{d} \psi}.
\end{equation*}

In this case the adjoint system is defined as
\begin{align*}
\boldsymbol{\lambda}_{n} - \boldsymbol{\lambda}_{n + 1}
&=
- \frac{ \partial j_{n + 1} }{ \partial \boldsymbol{\alpha}_{n + 1} }
+ \boldsymbol{\lambda}_{n + 1} \cdot 
\frac{ \partial \boldsymbol{\Delta}_{n + 1} }{ \partial \boldsymbol{\alpha}_{n + 1} }
\\
&=
- \mathbf{1} \cdot 
\frac{ \partial \boldsymbol{\Delta}_{n + 1} }{ \partial \boldsymbol{\alpha}_{n + 1} }
+ \boldsymbol{\lambda}_{n + 1} \cdot 
\frac{ \partial \boldsymbol{\Delta}_{n + 1} }{ \partial \boldsymbol{\alpha}_{n + 1} }
\\
&=
(\boldsymbol{\lambda}_{n + 1} - \mathbf{1}) \cdot 
\frac{ \partial \boldsymbol{\Delta}_{n + 1} }{ \partial \boldsymbol{\alpha}_{n + 1} }.
\end{align*}
The partial derivative reduces to
\begin{align*}
\frac{ \partial \Delta_{n, i} }{ \partial \alpha_{n, j} }
&=
\frac{ \partial }{ \partial \alpha_{n, j} } 
\left(
\omega_{n + 1, i} \, \sum_{k = 1}^{K} \Gamma_{n + 1, ik} \, \alpha_{n, k} - \alpha_{n, i}
\right)
\\
&=
\omega_{n + 1, i} \, \sum_{k = 1}^{K} \Gamma_{n + 1, ik} \, \delta_{jk} - \delta_{ij}
\\
&=
\omega_{n + 1, i} \, \Gamma_{n + 1, ij} - \delta_{ij}
\end{align*}
so that
\begin{equation*}
\sum_{i = 1}^{K} 
(\lambda_{n, i} - 1) \, \frac{ \partial \Delta_{n, i} }{ \partial \alpha_{n, j} }
=
\sum_{i = 1}^{K} 
(\lambda_{n, i} - 1) \, \omega_{n + 1, i} \, \Gamma_{n + 1, ij} - (\lambda_{n, j} - 1),
\end{equation*}
or in matrix notation,
\begin{equation*}
(\boldsymbol{\lambda}_{n} - \mathbf{1}) \cdot 
\frac{ \partial \boldsymbol{\Delta}_{n} }{ \partial \boldsymbol{\alpha}_{n} }
=
\boldsymbol{\Gamma}_{n + 1}^{\dagger} \cdot \left( \boldsymbol{\omega}_{n + 1}
\circ (\boldsymbol{\lambda}_{n} - \mathbf{1}) \right) - \boldsymbol{\lambda}_{n} + \mathbf{1}.
\end{equation*}
The backwards updates then become
\begin{align*}
\boldsymbol{\lambda}_{n} - \boldsymbol{\lambda}_{n + 1}
&=
(\boldsymbol{\lambda}_{n + 1} - \mathbf{1}) \cdot 
\frac{ \partial \boldsymbol{\Delta}_{n + 1} }{ \partial \boldsymbol{\alpha}_{n + 1} }
\\
\boldsymbol{\lambda}_{n} - \boldsymbol{\lambda}_{n + 1}
&=
\boldsymbol{\Gamma}_{n + 2}^{\dagger} \cdot \left( \boldsymbol{\omega}_{n + 2}
\circ (\boldsymbol{\lambda}_{n + 1} - \mathbf{1}) \right) - \boldsymbol{\lambda}_{n + 1} + \mathbf{1}
\\
\boldsymbol{\lambda}_{n}
&=
\boldsymbol{\Gamma}_{n + 2}^{\dagger} \cdot \left( \boldsymbol{\omega}_{n + 2}
\circ (\boldsymbol{\lambda}_{n + 1} - \mathbf{1}) \right) + \mathbf{1}.
\end{align*}
If we make the substitution 
\begin{equation*}
\boldsymbol{\kappa}_{n} = 1 - \boldsymbol{\lambda}_{n}
\end{equation*}
then this further simplifies to 
\begin{equation*}
\boldsymbol{\kappa}_{n}
=
\boldsymbol{\Gamma}_{n + 2}^{\dagger} \cdot \left( \boldsymbol{\omega}_{n + 2}
\circ \boldsymbol{\kappa}_{n + 1} \right),
\end{equation*}
which is just the backward states encountered above with a shifted index,
\begin{equation*}
\boldsymbol{\kappa}_{n} = \boldsymbol{\beta}_{n - 1}.
\end{equation*}

For the explicit derivative of the functional we also need
\begin{equation*}
\frac{ \partial j_{n} }{ \partial \psi}
-\boldsymbol{\lambda}_{n} \cdot 
\frac{ \partial \boldsymbol{\Delta}_{n} }{ \partial \psi}
=
(\mathbf{1} - \boldsymbol{\lambda}_{n} ) \cdot 
\frac{ \partial \boldsymbol{\Delta}_{n} }{ \partial \psi}
=
\boldsymbol{\kappa}_{n} \cdot 
\frac{ \partial \boldsymbol{\Delta}_{n} }{ \partial \psi},
\end{equation*}
where
\begin{equation*}
\frac{ \partial \boldsymbol{\Delta}_{n} }{ \partial \psi}
=
\frac{ \partial \boldsymbol{\omega}_{n + 1} }{ \partial \psi}
\circ ( \boldsymbol{\Gamma}_{n + 1} \cdot \boldsymbol{\alpha}_{n} )
+
\boldsymbol{\omega}_{n + 1}
\circ \left( \frac{ \partial \boldsymbol{\Gamma}_{n + 1} }{ \partial \psi} \cdot \boldsymbol{\alpha}_{n} \right).
\end{equation*}

Lastly we work out the boundary term. Recalling $\upsilon = \omega_0 \circ \rho$,
the boundary term is
\begin{align*}
\left[\mathbf{1} + \frac{ \partial j_{0} }{ \partial \boldsymbol{\alpha}_{0} }
- \boldsymbol{\lambda}_{0} \cdot \frac{ \partial \boldsymbol{\Delta}_{0} }{ \partial \boldsymbol{\alpha}_{0}  }
- \boldsymbol{\lambda}_{0} \right]^{\dagger} 
\cdot \frac{ \partial (\boldsymbol{\omega}_{0} \circ \rho)}{ \partial \psi }
&=
\left[ \mathbf{1} + \mathbf{1} \cdot \frac{ \partial \boldsymbol{\Delta}_{0} }{ \partial \boldsymbol{\alpha}_{0} }
- \boldsymbol{\lambda}_{0} \cdot \frac{ \partial \boldsymbol{\Delta}_{0} }{ \partial \boldsymbol{\alpha}_{0} }
- \boldsymbol{\lambda}_{0} \right]^{\dagger} 
\cdot \frac{ \partial (\boldsymbol{\omega}_{0} \circ \rho)}{ \partial \psi }
\\
&=
\left[ \left( \mathbf{1} - \boldsymbol{\lambda}_{0} \right) 
\cdot \frac{ \partial \boldsymbol{\Delta}_{0} }{ \partial \boldsymbol{\alpha}_{0} }
+ \mathbf{1} - \boldsymbol{\lambda}_{0} \right]^{\dagger}
\cdot \frac{ \partial (\boldsymbol{\omega}_{0} \circ \rho)}{ \partial \psi }
\\
&=
\Bigg[ \boldsymbol{\Gamma}_{1}^{\dagger} \cdot \left( \boldsymbol{\omega}_{1}
\circ ( \mathbf{1} - \boldsymbol{\lambda}_{0}  \right) - (\mathbf{1} - \boldsymbol{\lambda}_{0})
+ \mathbf{1} - \boldsymbol{\lambda}_{0} \Bigg]^{\dagger} 
\cdot \frac{ \partial (\boldsymbol{\omega}_{0} \circ \rho)}{ \partial \psi }
\\
&=
\Bigg[ \boldsymbol{\Gamma}_{1}^{\dagger} \cdot \left( \boldsymbol{\omega}_{1}
\circ ( \mathbf{1} - \boldsymbol{\lambda}_{0}  \right) \Bigg]^{\dagger} 
\cdot \frac{ \partial (\boldsymbol{\omega}_{0} \circ \rho)}{ \partial \psi }
\\
&=
\Bigg[ 
  \boldsymbol{\Gamma}_{1}^{\dagger} 
  \cdot \left( \boldsymbol{\omega}_{1} \circ \boldsymbol{\kappa}_{0} \right) 
\Bigg]^{\dagger} 
\cdot \frac{ \partial (\boldsymbol{\omega}_{0} \circ \rho)}{ \partial \psi }
\\
&=
\Bigg[ 
  \boldsymbol{\Gamma}_{1}^{\dagger} 
  \cdot \left( \boldsymbol{\omega}_{1} \circ \boldsymbol{\kappa}_{0} \right) 
\Bigg]^{\dagger} 
\cdot 
\left[ 
\boldsymbol{\omega}_{0} \circ \frac{ \partial \boldsymbol{\rho} }{ \partial \psi }
+ \frac{ \partial \boldsymbol{\omega}_{0} }{ \partial \psi } \circ \boldsymbol{\rho}
\right].
\end{align*}

Putting all of this together we can recover the derivative of the marginal likelihood
by computing
\begin{align*}
\frac{\mathrm{d}}{ \mathrm{d} \psi} \pi(y_{1}, \ldots, y_{N})
&= 
\mathbf{1}^{\dagger} \cdot \frac{ \mathrm{d} \boldsymbol{\alpha}_{N} }{ \mathrm{d} \psi } 
\\
&=
\left[\mathbf{1} + \frac{ \partial j_{0} }{ \partial \boldsymbol{\alpha}_{0} }
- \boldsymbol{\lambda}_{0} \cdot \frac{ \partial \boldsymbol{\Delta}_{0} }{ \partial \boldsymbol{\alpha}_{0}  }
- \boldsymbol{\lambda}_{0} \right]^{\dagger} 
\cdot \frac{ \partial (\boldsymbol{\omega}_{0} \circ \rho)}{ \partial \psi }
+ \sum_{n = 0}^{N - 1} 
\frac{ \partial j_{n} }{ \partial \psi}
-\boldsymbol{\lambda}_{n}^{\dagger} \cdot 
\frac{ \partial \boldsymbol{\Delta}_{n} }{ \partial \psi}
\\
&=
\quad 
\bigg[ 
\boldsymbol{\Gamma}_{1}^{\dagger} \cdot \left( \boldsymbol{\omega}_{1} \circ \boldsymbol{\kappa}_{0} \right) 
\bigg]^{\dagger}
\cdot 
\left[ 
\boldsymbol{\omega}_{0} \circ \frac{ \partial \boldsymbol{\rho} }{ \partial \psi }
+ \frac{ \partial \boldsymbol{\omega}_{0} }{ \partial \psi } \circ \boldsymbol{\rho}
\right]
\\
& \quad
+ \sum_{n = 0}^{N - 1} 
\boldsymbol{\kappa}_{n}^{\dagger} \cdot 
\left[ \frac{ \partial \boldsymbol{\omega}_{n + 1} }{ \partial \psi}
\circ \bigg( \boldsymbol{\Gamma}_{n + 1} \cdot \boldsymbol{\alpha}_{n} \bigg)
+
\boldsymbol{\omega}_{n + 1}
\circ \left( \frac{ \partial \boldsymbol{\Gamma}_{n + 1} }{ \partial \psi} \cdot \boldsymbol{\alpha}_{n} \right) \right],
\end{align*}
equivalent to the result from differentiating the expanded recursion.

One advantage to the discrete adjoint method is that we don't have to completely expand 
the recursion analytically, as done in the above derivation, or computationally, as would 
be done in a direct application of automatic differentiation.  Instead we can reason about 
the derivatives \emph{sequentially} in the same way that the system is originally defined.  

\pagebreak

\section{Conclusion}

In analogy to the continuous adjoint methods used with ordinary differential equations,
the discrete adjoint method defines a procedure to efficiently evaluate the derivatives 
of functionals over the evolution of discrete sequences. Because this procedure is fully 
defined by the derivatives of the forward difference equations and the summands defining 
the functional, it defines an efficient sequential differentiation algorithm that mirrors
the structure of the original sequence.  The beneficial scaling of this procedure makes 
the resulting implementations especially useful in practical applications.

We can apply the method to any mathematical model that depends on the parameters through 
an (implicit) forward difference equation.  Once we have made this equation explicit the 
derivation of a differentiation algorithm is completely mechanical, minimizing the burden
of its implementation.

\section*{Acknowledgements}

We thank Bob Carpenter for helpful discussions.

\bibliography{discrete_adjoints}
\bibliographystyle{imsart-nameyear}

\end{document}